\documentclass[conference]{IEEEtran}
\IEEEoverridecommandlockouts

\usepackage{cite}

\usepackage{color,xcolor}
\usepackage{epsfig}
\usepackage{graphicx}
\usepackage{lipsum}
\usepackage{array}
\usepackage{booktabs}
\usepackage{colortbl}
\usepackage{multirow}
\usepackage{float}
\usepackage{caption}
\usepackage{adjustbox}
\usepackage{subfigure}
\usepackage{footnote}
\makesavenoteenv{tabular}
\makesavenoteenv{table}
\usepackage{makecell}

\usepackage{amsmath,amsfonts,amssymb,bm}
\usepackage[super]{nth}

\usepackage{changepage}
\usepackage{extramarks}
\usepackage{lastpage}
\usepackage{setspace}
\usepackage{soul}
\usepackage{xspace}

\usepackage{url}
\usepackage{hyperref}
\hypersetup{colorlinks=True, urlcolor=black}

\usepackage{algorithm, algorithmic}
\usepackage{enumitem}
\usepackage{verbatim}
\usepackage{pifont}
\usepackage[acronyms]{glossaries}
\glsdisablehyper
  
\newcommand{\figdot}[1]{Fig.~\ref{fig:#1}}
\newcommand{\tbl}[1]{Table~\ref{tab:#1}}

\newcommand{\ignore}[1]{}

\makeatletter
\DeclareRobustCommand\onedot{\futurelet\@let@token\@onedot}
\def\@onedot{\ifx\@let@token.\else.\null\fi\xspace}

\makeatother

\definecolor{MyDarkBlue}{rgb}{0,0.08,1}
\definecolor{MyDarkGreen}{rgb}{0.02,0.6,0.02}
\definecolor{MyDarkRed}{rgb}{0.8,0.02,0.02}
\definecolor{MyDarkOrange}{rgb}{0.40,0.2,0.02}
\definecolor{MyPurple}{RGB}{111,0,255}
\definecolor{MyRed}{rgb}{1.0,0.0,0.0}
\definecolor{MyGold}{rgb}{0.75,0.6,0.12}
\definecolor{MyDarkgray}{rgb}{0.66, 0.66, 0.66}

\usepackage{pgfplots}
\DeclareUnicodeCharacter{2212}{−}
\usepgfplotslibrary{groupplots,dateplot}
\usetikzlibrary{patterns,shapes.arrows}
\pgfplotsset{compat=newest}

\begin{document}

\title{k2SSL: A Faster and Better Framework for Self-Supervised Speech Representation Learning}

% \author{Anonymous ICME submission}

\author{
\IEEEauthorblockN{Yifan Yang$^{1 \dagger}$, Jianheng Zhuo$^{1 \dagger}$, Zengrui Jin$^{3}$, Ziyang Ma$^{1}$, Xiaoyu Yang$^{2}$, Zengwei Yao$^{2}$ \\ Liyong Guo$^{2}$, Wei Kang$^{2}$, Fangjun Kuang$^{2}$, Long Lin$^{2}$, Daniel Povey$^{2}$, Xie Chen$^{1 *}$\thanks{$^{\dagger}$Equal contribution. $^{*}$Corresponding author.}}
\IEEEauthorblockA{
\textit{$^{1}$MoE Key Lab of Artificial Intelligence, X-LANCE Lab, Shanghai Jiao Tong University}\\
\textit{$^{2}$Xiaomi Corporation, Beijing, China}\\
\textit{$^{3}$The Chinese University of Hong Kong, Hong Kong SAR, China}\\
\{yifanyeung, zzasdf, chenxie95\}@sjtu.edu.cn
}
}

\maketitle

\begin{abstract}
Self-supervised learning (SSL) has achieved great success in speech-related tasks. While Transformer and Conformer architectures have dominated SSL backbones, encoders like Zipformer, which excel in automatic speech recognition (ASR), remain unexplored in SSL. Concurrently, inefficiencies in data processing within existing SSL training frameworks, such as fairseq, pose challenges in managing the growing volumes of training data. To address these issues, we propose k2SSL, an open-source framework that offers faster, more memory-efficient, and better-performing self-supervised speech representation learning, focusing on downstream ASR tasks. The optimized HuBERT and proposed Zipformer-based SSL systems exhibit substantial reductions in both training time and memory usage during SSL training. Experiments on LibriSpeech demonstrate that Zipformer Base significantly outperforms HuBERT and WavLM, achieving up to a 34.8\% relative WER reduction compared to HuBERT Base after fine-tuning, along with a 3.5x pre-training speedup in GPU hours. When scaled to 60k hours of LibriLight data, Zipformer Large exhibits remarkable efficiency, matching HuBERT Large's performance while requiring only 5/8 pre-training steps.
\end{abstract}

\section{Introduction}
Speech-based self-supervised learning (SSL)~\cite{wav2vec, wav2vec2, hubert, wavlm, data2vec, data2vec2, ssl} has emerged as a powerful paradigm for its generalization ability, leveraging vast amounts of unlabeled data to derive universal representations. These models can be easily fine-tuned by incorporating prediction layers for specific tasks, enhancing performance in applications like automatic speech recognition (ASR)~\cite{asr}. HuBERT~\cite{hubert}, one widespread self-supervised speech representation model, utilizes the classical and versatile Transformer~\cite{transformer} architecture and masked language modeling (MLM), achieving generalizability across various downstream tasks with remarkable performance \cite{superb}.

However, training speech-based SSL models demands enormous computational resources due to the need for massive quantities of unlabeled data, often more than ten times the amount of labeled data, for effective pre-training before fine-tuning.
For instance, training HuBERT Base requires 32 GPUs, taking 24 hours for the first iteration and 38 hours for the second iteration. When scaling up to larger variants, the requirement jumps to 128 GPUs for HuBERT Large and 256 GPUs for HuBERT X-Large. Similarly, WavLM~\cite{wavlm} training demands a lot of computational resources. WavLM Base needs 32 GPUs, while the WavLM Large requires 64 GPUs. The extensive memory usage and prolonged training time of these models make them inaccessible to most researchers, contradicting the fast-paced demands of the AI industry.

On the other hand, datasets used in both industry and academia have grown prohibitively large~\cite{basetts,voicebox,mms,whisper,usm,libriheavy,gigaspeech2,libriheavymix}, ranging from tens of thousands to hundreds of thousands of hours. These massive datasets have surpassed the capabilities of existing open-source speech-based SSL frameworks~\cite{faiseq,espnet,s3prl}, which struggle with excessive memory demands and inefficient data management when handling such large-scale data.
Consequently, there is considerable room for improving the accessibility, efficiency, and effectiveness of speech-based SSL systems.

With this perspective in mind, we propose k2SSL\footnote{Code and models are available at \url{https://github.com/k2-fsa/icefall}.}, an open-source framework designed to provide a faster, more memory-efficient, and higher-performing solution for self-supervised speech representation learning. k2SSL specifically addresses the inefficiencies encountered by existing frameworks when managing large-scale datasets, offering seamless scaling for both storage and training.

We optimize the HuBERT architecture by removing memory-intensive components that contribute little to performance, and we integrate the Zipformer~\cite{zipformer} encoder as the backbone, combined with the ScaledAdam~\cite{zipformer} optimizer. Zipformer Base pre-training requires only 8 V100 32G GPUs, while Zipformer Large requires 32 V100 32G GPUs, both without gradient accumulation, significantly reducing computational demands. This makes k2SSL more accessible and feasible for a broader range of research teams and organizations.
% 我们拿到了一些 insights
Extensive experiments on LibriSpeech~\cite{librispeech} and Libri-Light~\cite{librilight} demonstrate that Zipformer-based SSL systems exhibit strong performance and efficiency. Specifically, Zipformer Base achieves a relative WER reduction of up to 34.8\%/32.4\% on dev-other/test-other compared to HuBERT Base after supervised fine-tuning, while delivering a 3.5x pre-training speedup in total GPU hours, highlighting both efficiency and effectiveness. Zipformer Large demonstrates exceptional efficiency, achieving performance comparable to HuBERT Large while requiring only 5/8 of the pre-training steps.
\begin{figure*}[t]
    \centering
    \includegraphics[width=1\linewidth]{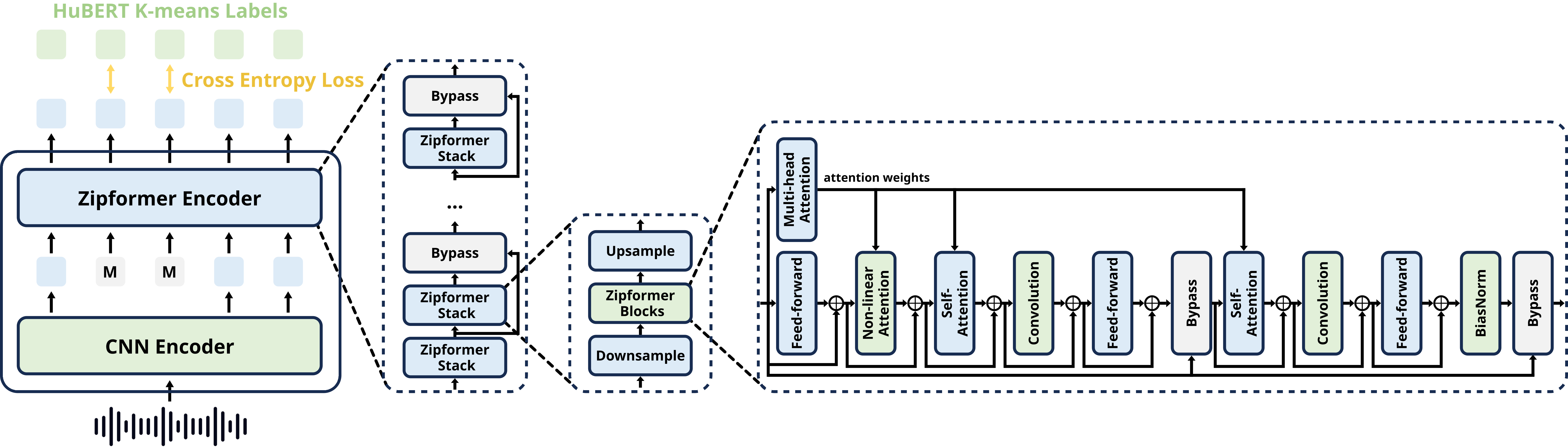}
    \vspace{0.5em}
    \caption{The overall architecture of the SSL system with Zipformer as the backbone.}
    \label{fig:system}
\end{figure*}

\section{HuBERT and its Variants}
\subsection{HuBERT}
HuBERT, short for Hidden-Unit BERT, is built on the wav2vec 2.0~\cite{wav2vec2} architecture and incorporates a convolutional feature extractor, a Transformer encoder, and a linear projection layer. It processes raw speech and predicts $k$-means clustering labels for masked audio segments. HuBERT processes audio at a standard 16kHz rate, downsampling the sequences to 50Hz through the convolutional feature extractor.
A predetermined percentage of timestamps are selected as starting points for mask spans, covering multiple consecutive timestamps to form the masked sections of the sequence. For these masked parts, the speech features are replaced with a learnable mask embedding. The model then calculates the probability distribution across cluster labels using the following equation:
\begin{equation}
    p_{f}(c | \tilde{X}, t)=\frac{\exp \left(\text{sim}\left(A o_{t}, e_{c}\right) / \tau\right)}{\sum_{i=1}^{C} \exp \left(\text{sim}\left(A o_{t}, e_{i}\right) / \tau\right)}
    \label{eqn:hubert_loss}
\end{equation}
Here, $\tilde{X}$ represents the masked input, $[o_1, ..., o_T]$ the output from the Transformer encoder, $A$ the projection matrix, $e_c$ the cluster label embedding, $C$ the number of clusters, and $\tau$ the logit scaler, set to a specific value.
HuBERT Base is pre-trained in two iterations. The first iteration uses 39-dimensional MFCC features from the LibriSpeech-960h dataset, clustered with k-means using a codebook size of 100. In the second iteration, features from the 6th layer of the first model are clustered with a codebook size of 500. HuBERT Large is pre-trained on clusters derived from the 9th layer of the second iteration, again using a codebook size of 500.

Several follow-up works~\cite{cmu_hubert,melhubert,fasthubert,hubert-ap,ssl_discrete,ctcbert,pbert,polybert,wavlm} have been proposed to boost either efficiency or effectiveness based on HuBERT. 

\subsection{Efficient Variants}
MelHuBERT~\cite{melhubert} simplifies HuBERT training by using Fbank as input and replacing the pre-train loss with the cross-entropy.
Fast-HuBERT~\cite{fasthubert} also accelerates HuBERT training by utilizing Fbank features with larger frameshifts as input. However, using larger frameshift Fbank features requires extracting them in advance, which introduces extra time and storage costs and leads to a notable decline in performance.

\subsection{Effective Variants}
HuBERT-AP~\cite{hubert-ap} and CTCBERT~\cite{ctcbert} change how the Transformer output is aligned with the SSL units, making the pre-training process more like the common practice of speech recognition. 
Nonetheless, these implementation modifications have limited performance improvement over the original HuBERT. 
PBERT~\cite{pbert}, MonoBERT~\cite{polybert}, and PolyBERT~\cite{polybert} introduce phoneme information to the SSL units to improve performance. However, a pre-trained phoneme recognizer with pair data is needed in PBERT, and additional wav2vec-U 2.0~\cite{wav2vec-u2} is required in MonoBERT and PolyBERT. 
WavLM~\cite{wavlm} effectively enhances the multi-task performance by increasing the data scale and incorporating additional denoising loss during the pre-training stage, but it requires a substantial computational cost.
Boosting efficiency and effectiveness simultaneously in speech-based SSL models remains to be explored.
\section{k2SSL}
k2SSL aims to enhance both the efficiency and effectiveness of HuBERT-based SSL systems. We begin by analyzing design inefficiencies in these systems, which lead to unnecessary computational overhead, and present our corrective modifications. We then describe our transition from the Transformer encoder to Zipformer and the integration of the ScaledAdam optimizer. Finally, we outline several engineering advantages of the proposed k2SSL framework. The overall SSL system, with Zipformer as the backbone, is depicted in \figdot{system}.

\subsection{Analysis and Optimization of HuBERT Architecture}
HuBERT incorporates the convolutional feature extractor from the wav2vec 2.0 architecture to process audio sampled at 16kHz, which includes 7-layer convolutional layers. In HuBERT Base, an Fp32 GroupNorm is applied after the first convolution, while in HuBERT Large, an Fp32 LayerNorm is applied after each convolution. Testing on an idle V100 GPU with a batch of six 60-second audio samples showed peak memory usage for the convolutional feature extractor: 10.96 GB for HuBERT Base and 13.14 GB for HuBERT Large, reflecting a 19.9\% increase in memory overhead.
Our experiments indicate that although applying Fp32 LayerNorm after each layer enhances stability during half-precision training, it contributes little to performance and incurs huge memory overhead. Instead, we retain a single Fp32 GroupNorm in the first layer and resort to the half-precision stable Zipformer + ScaledAdam framework.

Following \cite{melhubert}, cross-entropy loss is used for pre-training, largely reducing peak memory usage by eliminating updates to codeword embeddings and cosine similarity calculations. This simplified loss function predicts the $k$-means clustering labels of the masked input $\tilde{X}$ as follows:
\begin{equation}
    p_{f}(c | \tilde{X}, t)=\frac{\exp \left((A o_t)_c\right)}{\sum_{i=1}^{C} \exp \left((A o_t)_i\right)},
    \label{eqn:simplified_hubert_loss}
\end{equation}
where $o_t$ denotes the Transformer encoder output at time step $t$, and $A$ represents the projection matrix.

\subsection{Adapting Zipformer and ScaledAdam for SSL}
The Transformer encoder exhibits quadratic memory and computational complexity as the input sequence length increases. However, self-supervised learning benefits more from longer sequences, creating a conflict between efficiency and performance. Furthermore, as noted in \cite{hubert}, the total batch size is a critical factor influencing model performance.

To tackle these, we transition to Zipformer, which handles long sequences more memory efficiently and supports larger batch sizes, enabling more efficient and effective SSL pre-training.
As shown in \figdot{system}, Zipformer~\cite{zipformer} encoder adopts a U-Net-like structure to learn temporal representations efficiently at varying resolutions across different stacks. Starting with an acoustic feature sequence at a 50Hz frame rate, Zipformer processes it through six cascaded stacks, with frame rates of 50Hz, 25Hz, 12.5Hz, 6.25Hz, 12.5Hz, and 25Hz. Except for the first stack, each downsamples the sequence, processes it, and then upsamples it back to 50Hz.
The embedding dimensions vary across the stacks, with the central stacks having larger dimensions. The output from each stack is truncated or padded with zeros to match the dimensions of the subsequent stack. The final encoder output is set to the maximum dimension across all stacks, concatenating segments from each stack's output and taking the most recent output for each dimension. Zipformer also incorporates several upgrades: a redesigned block structure that reuses attention weights for efficiency, BiasNorm~\cite{zipformer} for better retention of sequence length, and activation functions SwooshR~\cite{zipformer} and SwooshL~\cite{zipformer}, surpassing the performance of Swish.
Additionally, ScaledAdam~\cite{zipformer}, a parameter-scale-invariant version of Adam, decouples gradient scale and direction, further improving convergence.

\begin{table}[t]
  \centering
  \caption{Hyperparameters of various models in k2SSL for pre-training and fine-tuning. Base models are pre-trained on 8 NVIDIA V100 32GB GPUs, and Large models are on 32 NVIDIA V100 32GB GPUs. All models are fine-tuning on 8 NVIDIA V100 32GB GPUs. ``LR'' represents the learning rate, ``BS'' represents the duration of speech samples in a single batch, and ``GA'' represents gradient accumulation.}
  \label{tab:setup}
  \renewcommand{\arraystretch}{1.2}
  \renewcommand\tabcolsep{2.0pt}
  \resizebox{\linewidth}{!}{
    \begin{tabular}{cccccccc}
    \toprule[1.5pt]
    \multirow{2}{*}{\textbf{Backbone}} & \multicolumn{2}{c}{{\bf Stage}} & \multirow{2}{*}{\textbf{Loss}} & \multirow{2}{*}{\textbf{\# Params}} & \multirow{2}{*}{\textbf{LR}} & \multirow{2}{*}{\textbf{BS/GPU}} & \multirow{2}{*}{\textbf{GA}} \\
    & {\bf Pre-train} & {\bf Fine-tune} &&&&& \\
    \midrule
    \multirow{2}{*}{Transformer} & \checkmark & & HuBERT       & 94.7M  & 0.045   & 87.5s  & 4 \\
                                 & & \checkmark & Pruned RNN-T & 96.5M  & 0.001   & 200.0s & 1 \\
    \midrule
    \multirow{2}{*}{Transformer} & \checkmark & & CE           & 94.8M  & 0.045   & 87.5s  & 4 \\
                                 & & \checkmark & Pruned RNN-T & 96.6M  & 0.00075 & 200.0s & 1 \\
    \midrule
    \multirow{3}{*}{Zipformer}   & \checkmark & & CE           & 94.6M  & 0.045   & 600.0s & 1 \\ 
                                 & & \checkmark & Pruned RNN-T & 96.4M  & 0.002   & 600.0s & 1 \\
                                 & & \checkmark & CTC          & 94.6M  & 0.001   & 600.0s & 1 \\
    \midrule
    \multirow{2}{*}{Zipformer}   & \checkmark & & CE           & 305.9M & 0.045   & 350.0s & 1 \\ 
                                 & & \checkmark & Pruned RNN-T & 308.0M & 0.002   & 400.0s & 1 \\
    \bottomrule[1.5pt]
    \end{tabular}
  }
\end{table}
\begin{table*}[t]
  \centering
  \caption{Comparison of Word Error Rates (WERs) \textbf{without an external language model} during decoding on LibriSpeech dev/test sets and pre-training speedup ratios measured in total GPU hours. \textbf{All models are pre-trained on unlabeled LibriSpeech-960h and fine-tuned on labeled LibriSpeech-100h using pruned RNN-T loss, 500-class BPE word pieces as modeling units}, and ScaledAdam as optimizer.}
  \label{tab:result_ls100_rnnt}
  \renewcommand{\arraystretch}{1.2}
  \renewcommand\tabcolsep{3.0pt}
  \resizebox{\linewidth}{!}{
  \begin{tabular}{ccccccccccc}
    \toprule[1.5pt]
    \multirow{2}{*}{\textbf{Model}} & \multirow{2}{*}{\textbf{Backbone}} & \multicolumn{3}{c}{\textbf{Pre-train}} & \multicolumn{4}{c}{{\bf Word Error Rate (\%)}} & \multirow{2}{*}{\makecell[c]{\textbf{GPU} \\ \textbf{Hours}}} & \multirow{2}{*}{\makecell[c]{\textbf{Pre-train} \\ \textbf{Speedup}}} \\ 
    && \textbf{Toolkit} & \textbf{Optimizer} & \textbf{Loss Function} & \textbf{dev-clean} & \textbf{dev-other} & \textbf{test-clean} & \textbf{test-other}\\
    \midrule
    HuBERT Base    & Transformer & fairseq & Adam & HuBERT
                   & 4.88 & 12.06 & 4.98 & 11.65 & 1878 & 1x \\
    \midrule
    HuBERT Base    & Transformer & k2SSL & ScaledAdam & HuBERT
                   & 4.67$_{\textcolor{teal}{-4.3\%}}$ & 12.05$_{\textcolor{teal}{-0.0\%}}$ & 4.81$_{\textcolor{teal}{-3.4\%}}$ & 11.53$_{\textcolor{teal}{-1.0\%}}$ & 1270 & 1.48x \\
    HuBERT Base    & Transformer & k2SSL & ScaledAdam & CE
                   & 4.54$_{\textcolor{teal}{-7.0\%}}$ & 11.61$_{\textcolor{teal}{-3.7\%}}$ & 4.77$_{\textcolor{teal}{-4.2\%}}$ & 11.09$_{\textcolor{teal}{-4.8\%}}$ & 886 & 2.12x \\
    Zipformer Base & Zipformer   & k2SSL & ScaledAdam & CE
                   & \textbf{3.67}$_{\textcolor{teal}{-24.8\%}}$ & \textbf{7.86}$_{\textcolor{teal}{-34.8\%}}$  & \textbf{3.80}$_{\textcolor{teal}{-23.7\%}}$ & \textbf{7.87}$_{\textcolor{teal}{-32.4\%}}$ & 531 & \textbf{3.53x} \\
    \bottomrule[1.5pt]
    \end{tabular}
    }
\end{table*}

\subsection{Fine-tuning with Pruned RNN-T Loss}
Letter-level CTC~\cite{ctc} loss is commonly employed for downstream ASR in speech-based SSL, but it is inefficient, slow to converge, and hard to align.
In addition to using letter-level CTC loss for fair comparisons, we leverage pruned RNN-T loss~\cite{pruned-rnnt}, a memory-efficient variant of transducer loss~\cite{rnn-t}, to enhance ASR fine-tuning effectiveness with minimal overhead.

\subsection{Technical Advantages of k2SSL}
k2SSL integrates Lhotse~\cite{lhotse} to efficiently manage large-scale datasets by optimizing I/O bandwidth and storage capacity, enabling seamless scaling for both storage and training. It excels in processing speech segments stored in shuffled order by duration, utilizing the \texttt{DynamicBucketingSampler} from Lhotse to eliminate the need to load entire manifests into memory. By streamingly reading a subset of manifests to estimate bucket boundaries and maintaining a fixed-size buffer, this approach ensures consistent data loading times, regardless of dataset size.
As a result, pre-training on vast datasets like the 60,000-hour Libri-Light~\cite{librilight} is initiated in merely dozens of seconds, in contrast to the lengthy loading times and huge memory demands of traditional frameworks.

Moreover, pre-trained HuBERT checkpoints from fairseq\footnote{\url{https://github.com/facebookresearch/fairseq/tree/main/examples/hubert}} are compatible with k2SSL, supporting further fine-tuning for downstream tasks.
\section{Experiments}
\label{sec:exp}
\subsection{Experimental Setups}
Four distinct experimental setups are detailed in \tbl{setup}, with all Base models containing ($\sim$95M) parameters and Large model containing ($\sim$306M) parameters.
Setups 1 and 2 (Rows 1 \& 2) utilize the Transformer encoder as the backbone. Setup 1 uses the original HuBERT loss, and Setup 2 adopts cross-entropy (CE) loss during pre-training. Setups 3 and 4 (Rows 3 \& 4) employ Zipformer Base and Zipformer Large with CE loss for pre-training.
Setups 1, 3, and 4 set batch sizes to fully exploit single GPU memory. With the memory-efficient designs, Zipformer Base and Zipformer Large require only 8 and 32 GPUs, respectively, without gradient accumulation. This allows for a larger effective batch size while requiring only 1/4 of the GPUs compared to the original HuBERT~\cite{hubert}, which necessitates 32 and 128 GPUs.

\noindent\textbf{Dataset}\quad
960 hours of LibriSpeech~\cite{librispeech} audio or 60,000 hours of Libri-light~\cite{librilight} audio are used for pre-training. 100 hours or 960 hours of LibriSpeech are considered for fine-tuning.

\noindent\textbf{Pre-training}\quad
We directly utilize the target labels identical to those used in HuBERT~\cite{hubert} and WavLM~\cite{wavlm} for fair comparisons. Since these targets are well-established, there is no need for metrics like phone purity, cluster purity, or PNMI, aligning with our intention to attribute performance differences to model and training variations rather than target quality.
All Base models are pre-trained for 300 epochs (400k steps for HuBERT Base, 225k steps for Zipformer Base) on unlabeled 960 hours of LibriSpeech audio, using 500-class k-means labels\footnote{\href{https://huggingface.co/datasets/yfyeung/librispeech-hubert-base-ls960-iter1-l6-km500}{librispeech-hubert-base-ls960-iter1-l6-km500}} derived from clustering the 6th layer of the first iteration HuBERT Base model.
Zipformer Large is pre-trained for 250k steps on 60,000 hours of Libri-Light, using 500-class k-means labels\footnote{\href{https://huggingface.co/datasets/yfyeung/librilight-hubert-base-ls960-iter2-l9-km500}{librilight-hubert-base-ls960-iter2-l9-km500}} extracted from the 9th layer of the second iteration HuBERT Base model\footnote{\url{https://dl.fbaipublicfiles.com/hubert/hubert_base_ls960.pt}}. The same masking strategies as in \cite{hubert} are used. ScaledAdam with $\beta = (0.9, 0.98)$ is applied alongside Eden~\cite{zipformer} scheduler.

\noindent\textbf{Fine-tuning with Pruned RNN-T Loss}\quad
500-class Byte Pair Encoding (BPE)~\cite{bpe} word pieces are selected as the classification units. 
For decoding, we adopt constrained beam search~\cite{fast_decoding} with a beam size of 8, which permits only one symbol emission per frame.

\noindent\textbf{Fine-tuning with CTC Loss}\quad
The CTC vocabulary comprises 26 English letters, a space, an apostrophe, and a CTC blank symbol.
The decoding process with an external language model follows \cite{hubert}, utilizing the wav2letter++~\cite{wav2letter++} beam search decoder, formulated as follows:
\begin{equation}
    \log p_{CTC}(\boldsymbol{y} \mid \boldsymbol{x})+w_{1} \log p_{LM}(\boldsymbol{y})+w_{2}|\boldsymbol{y}|,
\end{equation}
where $\boldsymbol{x}$ is the input audio, $\boldsymbol{y}$ is the predicted text, $|\boldsymbol{y}|$ is length of the text, and $w_1$ and $w_2$ are the language model weight and word score coefficients.

\begin{table}[t]
  \centering
  \caption{Comparison of WERs w/wo a language model between Zipformer Base and baseline models on LibriSpeech dev/test sets. \textbf{All models are pre-trained on unlabeled LibriSpeech-960h and fine-tuned on labeled LibriSpeech-100h, using CTC loss and letters as modeling units}. The results of baseline models are collected from their respective papers or fine-tuned based on official checkpoints.}
  \label{tab:result_ls100}
  \renewcommand{\arraystretch}{1.2}
  \renewcommand\tabcolsep{2.0pt}
  \resizebox{\linewidth}{!}{
  \begin{tabular}{cccccc}
    \toprule[1.5pt]
    \multirow{2}{*}{\textbf{Model}} & \multirow{2}{*}{\textbf{LM}} & \multicolumn{4}{c}{{\bf Word Error Rate (\%)}} \\ 
    && \textbf{dev-clean} & \textbf{dev-other} & \textbf{test-clean} & \textbf{test-other} \\
    \midrule
    wav2vec 2.0 Base~\cite{wav2vec2} & None   & 6.1  & 13.5 & 6.1  & 13.3 \\
    HuBERT Base                      & None   & 5.3  & 13.0 & 5.4  & 12.6 \\
    WavLM Base~\cite{wavlm}          & None   & -    & -    & 5.7  & 12.0 \\
    Zipformer Base                   & None   & \textbf{4.7} & \textbf{9.6} & \textbf{4.4} & \textbf{9.8} \\
    \midrule
    wav2vec 2.0 Base~\cite{wav2vec2} & 4-gram & 2.7  & 7.9  & 3.4  & 8.0 \\
    HuBERT Base~\cite{hubert}        & 4-gram & 2.7  & 7.8  & 3.4  & 8.1 \\
    WavLM Base~\cite{wavlm}          & 4-gram & -    & -    & 3.4  & 7.7 \\
    Zipformer Base                   & 4-gram & \textbf{2.6} & \textbf{6.4} & \textbf{3.0}  & \textbf{6.8} \\
    \bottomrule[1.5pt]
    \end{tabular}
    }
\end{table}
\begin{table*}[t]
  \centering
  \caption{Comparison of WERs \textbf{without an external language model} on LibriSpeech test sets. All models are \textbf{trained with transducer or pruned RNN-T loss on labeled LibriSpeech-960h}. All pre-trained models are compared based on the number of pre-training steps required to achieve similar performance, rather than maximizing performance.}
  \label{tab:result_ls960_rnnt}
  \renewcommand{\arraystretch}{1.2}
  \renewcommand\tabcolsep{10pt}
  \resizebox{0.85\linewidth}{!}{
  \begin{tabular}{lcccc}
    \toprule[1.5pt]
    \multirow{2}{*}{\textbf{Model}} & \multirow{2}{*}{\textbf{Unlabeled Data}} & \multirow{2}{*}{{\bf Pre-train Steps}} & \multicolumn{2}{c}{{\bf Word Error Rate (\%)}} \\
    & & & \textbf{test-clean} & \textbf{test-other} \\
    \midrule
    \textbf{Supervised} \\
    \hspace{1em} Transformer Transducer~\cite{transformer_transducer} & - & - &  2.4 & 5.6 \\
    \hspace{1em} Conformer-L Transducer~\cite{conformer}              & - & - & 2.1 & 4.3 \\
    \hspace{1em} Conformer-L Pruned Transducer~\cite{zipformer}       & - & - & 2.5 & 5.6 \\ 
    \hspace{1em} Zipformer-L Pruned Transducer~\cite{zipformer}       & - & - & 2.1 & 4.6 \\
    \midrule
    \textbf{Pre-trained} \\
    \hspace{1em} Conformer-L~\cite{conformer_pretrain_rnnt} & LL-60k  & 400k & 2.0 & 4.5 \\
    \hspace{1em} HuBERT Large                               & LL-60k  & 400k & \textbf{1.8} & \textbf{3.9} \\
    \hspace{1em} Zipformer Large                            & LL-60k  & \textbf{250k} & \textbf{1.8} & 4.0 \\
    \bottomrule[1.5pt]
    \end{tabular}
    }
\end{table*}

\subsection{Experimental Results}
\tbl{result_ls100_rnnt} presents the ASR performance of the fairseq pre-trained HuBERT Base (Row 1), k2SSL pre-trained HuBERT Base with two distinct pre-training losses (Rows 2 \& 3), and k2SSL pre-trained Zipformer Base (Row 4). All models are fine-tuned within the k2SSL framework on the LibriSpeech-100h, using the pruned RNN-T loss. The following conclusions can be drawn from the analysis:
\begin{itemize}
\item Compared to the fairseq pre-trained HuBERT Base, the k2SSL pre-trained HuBERT Base with the assistance of ScaledAdam optimizer achieves a speedup ratio of approximately 1.5 (Row 2 \textit{vs.} 1); 
\item The simplified pre-training loss further provides a around 2.1x pre-training speedup and a relative WER reduction of up to 3.8\% (Row 3 \textit{vs.} 2, test-other);
\item When using Zipformer as the backbone, the largest WER reduction is obtained across all dev and test sets of LibriSpeech (Row 4 \textit{vs.} 3). Compared to the original HuBERT Base, Zipformer Base achieves significant relative WER reductions by up to 24.8\% on dev-clean, 34.8\% on the dev-other, 23.7\% on test-clean, and 32.4\% on test-other while yielding a 3.53 pre-training speedup (Row 4 \textit{vs.} 1). This indicates the superiority of the k2SSL framework with the Zipformer encoder as the backbone.
\end{itemize}

\tbl{result_ls100} presents the ASR performance of Zipformer Base against three leading SSL models. The results are either collected from the original paper or fine-tuned on LibriSpeech-100h using CTC loss with letter-level modeling units based on publicly available checkpoints. The key observations from the results can be made:
\begin{itemize}
    \item Without language model fusion, Zipformer Base significantly outperforms wav2vec 2.0 Base, HuBERT Base, and WavLM Base by relative WER reductions up to 28.9\% (Row 4 \textit{vs.} 1, dev-other), 26.2\% (Row 4 \textit{vs.} 2, dev-other), and 22.8\% (Row 4 \textit{vs.} 3, test-clean), respectively;
    \item The performance advancement of Zipformer Base is maintained when incorporating an external 4-gram language model for decoding. Zipformer Base brings a relative WER reduction of up to 19.0\% (Row 8 \textit{vs.} 5, dev-other), 18.0\% (Row 8 \textit{vs.} 6, dev-other), and 11.8\% (Row 8 \textit{vs.} 7, test-clean), respectively.
\end{itemize}

\tbl{result_ls960_rnnt} reports the ASR performance of Large variants on the full LibriSpeech. Results are either sourced from the original paper or fine-tuned on LibriSpeech-960h within the k2SSL framework, using pruned RNN-T loss with publicly available checkpoints. The main takeaways from the results are:
\begin{itemize}
    \item Both HuBERT Large and Zipformer Large, fine-tuned within the k2SSL framework, outperform all supervised models;
    \item Zipformer Large (308.0M) achieves comparable performance against HuBERT Large (318.7M) with much fewer pre-training steps (250k \textit{vs.} 400k). We are optimistic that more pre-training steps will yield better results, as observed in the Base variant.
\end{itemize}

These observations show the remarkable capabilities and advancements of the k2SSL approach in improving both efficiency and effectiveness.
\section{Conclusions}
This paper introduces k2SSL, a scalable framework designed for more efficient and effective self-supervised speech representation learning, tailored for downstream ASR tasks. By optimizing the HuBERT architecture and adopting Zipformer as the backbone, k2SSL achieves remarkable reductions in both memory usage and training time, making it accessible to a broader range of research teams and organizations. Extensive experiments on LibriSpeech and Libri-Light validate the superiority of k2SSL, with Zipformer-based SSL systems significantly outperforming HuBERT and WavLM, achieving a 3.5x pre-training speedup and notable WER reductions. The related resources, including the k2SSL framework, training configurations, and pre-trained models, are publicly available to facilitate further research.
\section*{Acknowledgment}
This work was supported by the National Natural Science Foundation of China  (No. U23B2018 and No. 62206171), Shanghai Municipal Science and Technology Major Project under Grant 2021SHZDZX0102 and the International Cooperation Project of PCL.

\bibliographystyle{IEEEtran}
\bibliography{refs}

\end{document}